\def\ref{\hangindent=20pt \hangafter=1 \noindent}
\documentstyle[12pt]{article}   
\begin{document}    
\baselineskip=24.5pt    
\setcounter{page}{1}         
\topskip 0 cm    
\vspace{1 cm}    
\centerline{\Large \bf Lepton Flavour Mixing Matrix and CP Violation}
\centerline{\Large \bf from Neutrino Oscillation Experiments}
\vskip 1 cm
\centerline{\large M. Fukugita$^{1}$ and M. Tanimoto$^2$}
\vskip5mm
\centerline{$^1$ Institute for Cosmic Ray Research, University of Tokyo,
Kashiwa 2778582, Japan}
\centerline{$^2$ Department of Physics, Niigata University, Niigata 950-2181,
Japan}

%\centerline{04  July 2001}
 
\vskip 4 cm
\noindent
{\large\bf Abstract}
\medskip

\noindent
The measurement of the charged-current $^8$B solar neutrino reaction
on deuterium at the Sudbury Neutrino Observatory confirms the neutrino
oscillation hypothesis for the solar neutrino problem, and the result
favours the solution with large neutrino mixing angles. We demonstrate
that the current neutrino oscillation data (including atmospheric and
reactor neutrinos) are sufficient to construct
the lepton flavour mixing matrix with a reasonable accuracy. We also
infer the maximum size of CP violation effects consistent with the
current neutrino oscillation experiments.

\newpage 

\noindent
The Sudbury Neutrino Observatory (SNO) experiment [1] measured the charged
current rate for $^8$B solar neutrino reactions on deuterium. The
derived electron neutrino flux is lower by $3.3\sigma$ than the
neutrino flux obtained from the electron scattering event rate measured
by the same experiment, and more accurately by Super-Kamiokande (SK) [2]. This
difference is ascribed to the neutral-current induced reaction of
three species of the left-handed neutrinos on electron scattering.
The sum of the 
electron neutrino flux and that of other neutrinos, 
inferred from the difference,
agrees with the electron neutrino generation rate expected
in the sun [3], hereby confirming the neutrino oscillation hypothesis
as the origin of the long-standing solar neutrino problem.    

The SNO result also narrows the range of the neutrino oscillation
parameters. The rate of the conversion of electron neutrinos to
other neutrinos is somewhat larger than is expected in the optimal
solutions within the neutrino oscillation hypothesis (e.g., [2],[4])
inferred from solar neutrino experiments  available 
prior to the SNO experiment [5],[2],[6]. 
This indicates that the new optimal solution
including the SNO data is located slightly inwards within the MSW [7]
triangle. This makes the small angle solution (SMA) of the MSW effect 
disfavoured, since the absence of the 
distortion of the neutrino
energy spectrum and of day-night effect observed at SK leaves
only the small angle edge of the SMA to be allowed [2], 
contrary to the charged-current event at SNO indicates. 
For a similar reason the neutrino oscillation
hypothesis {\it in vacuo} is now disfavoured. We are left with 
the large mixing angle
solution (LMA) and the LOW solution [4] 
of the MSW conversion 
as the most likely solution to the solar neutrino problem.
Detailed statistical analyses including the SNO data have already been 
made [8],[9],[10], and confirmed the picture we have sketched here.

In this {\it Letter} we present an analysis of the neutrino oscillation
 from a somewhat different view. We ask the question whether we can determine
the lepton flavour mixing matrix using the presently available oscillation 
data, and answer this question positively. We also ask the magnitude of
the CP violation effect allowed by the current neutrino oscillation
experiment.

We write the neutrino mixing matrix 

\begin{equation}
|\psi_\alpha\rangle = U_{\alpha i}|\psi_i\rangle ,
\end{equation}  
where $\alpha=e,\mu,\tau$ and $i=1,2,3$. The solar neutrino experiments
tell us about $|U_{e1}|$ and $|U_{e2}|$, and the atmospheric [11],[12] and
the K2K long base-line accelerator experiment [13] tell us 
about $|U_{\mu3}|$. Another constraint is imposed on the element $|U_{e3}|$
 from the Chooz reactor experiment [14]. 
If errors are small this information should be 
sufficient to construct the full $3\times3$ lepton flavour
mixing matrix (up to the phase) under the
unitarity constraint on the matrix. 

We consider
all constraints at a 90\% confidence level. We fix for simplicity the 
mass square difference between $\nu_\tau$ and $\nu_\mu$ to be
$3.2\times 10^{-3}$ (eV)$^2$ as deduced from the atmospheric 
neutrino experiment of SK [11]. 
We determine allowed regions of the Kobayashi-Maskawa
angles from the oscillation parameters and mapped them into
physical mixing matrix elements.
The present experimental information is not
sufficient to constrain the phase factor. So we vary
the phase $\phi$ between 0 and $\pi$ while searching for the
allowed angles, adopting the matrix representation 
in {\it Review of Particle Physics} [15].  

We take LMA for the solar neutrino mixing solution. 
The mixing matrix we derived reads

\begin{equation}
U= \left[ \matrix{0.74-0.90 & 0.45-0.65 & <0.16 \cr
                  0.22-0.61 & 0.46-0.77 & 0.57-1/\sqrt 2 \cr
                  0.14-0.55 & 0.36-0.68 & 1/\sqrt 2-0.82 \cr
                                         } \right]\ ,
\end{equation}
% New
%\begin{equation}
%U= \left[ \matrix{1/\sqrt 2-0.89 & 0.45-1/\sqrt 2 & <0.2 \cr
%                  0.19-0.66 & 0.41-0.78 & 0.57-1/\sqrt 2 \cr
%                  0.12-0.60 & 0.29-0.70 & 1/\sqrt 2-0.82 \cr
%                                         } \right]
%\end{equation}
where we are confined to the case of $\nu_\mu<\nu_\tau$, and 
only the modulus of the elements are shown.
If we allow for $\nu_\mu>\nu_\tau$ the
(2,3) element takes $1/\sqrt 2 - 0.82$ and the (3,3) element is
$0.57 -  1/\sqrt 2$, {\it i.e.}, the two elements are interchanged as 
the phase is unconstrained in our analysis.
It is interesting to
note that {\it all} matrix elements are reasonably constrained
with the present neutrino oscillation data. This predicts the
oscillation properties between any kinds of neutrinos. It is also
interesting to notice that all elements except for $U_{e3}$
are sizable. A marginal disparity is seen 
between the $U_{e3}$ and $U_{\tau1}$ elements, but more accurate
input of $U_{e2}$ is needed for a definitive conclusion.
%the latter is at least by a factor of two larger than the former. 
The small $U_{e3}$
is the characteristic that has been predicted in some 
phenomenological neutrino mass matrix models [16],[17] prior to
the Chooz experiment\footnote{Note 
that (1,3) and (3,1) elements are reversely expressed in [16]
by convention.}. 
%%%%%%%%%%%%%%%%%%%%%%%%%%%%%%%%%%%%%%%%%%%%%%%%%%%

Once this matrix is determined one can infer the maximum size of  
CP violation in the lepton sector. In the last few years the 
feasibility of detecting CP violation has been studied by a number
of authors in view of long-baseline neutrino experiments with strong
neutrino beams [18],[19]. In these studies only a few representative 
values of neutrino oscillation parameters, as written in terms of 
the angle
representation, are adopted to examine experimental feasibility, and 
systematic parameter searches are not made. 
A more general analysis is straightforward
with our neutrino matrix. The factor that represents the net 
CP violation effect can be written as [20]:

\begin{equation}
J={|U_{e1}|~|U_{e2}|~|U_{\mu3}|~|U_{\tau3}|~|U_{e3}| \over
 1-|U_{e3}|^2} \sin\phi\ .
\end{equation}
This is evaluated using the lepton mixing matrix, but more conveniently
expressed only with experimentally relevant quantities, as

\begin{equation}
J={1\over 4}{\sqrt{\sin^22\theta_{sol}}\sqrt{\sin^22\theta_{atm}}|U_{e3}| \over
 1-|U_{e3}|^2} \sin\phi\ ,
\end{equation}
where $\theta_{sol}$ is the mixing angle that directly comes into solar
neutrino oscillation, $\theta_{atm}$ is that for the atmospheric
neutrino oscillation, and $|U_{e3}|$ is directly constrained by the
$\bar\nu_e\rightarrow \bar\nu_\tau$ oscillation experiment.
Now it is easy to show that the maximum value of $J$ is given by

\begin{equation}
%J\leq {1\over 4}{\sqrt{0.93}\  \sqrt{1}\  0.16\over 1-0.16^2}\sin\phi
J\leq 0.040 \sin\phi\ .
\label{eq:5}
\end{equation}

Apart from the phase factor $\sin\phi$ the crucial factor that controls
the feasibility of the CP violation experiment is $|U_{e3}|$. This 
demonstrates the importance of the measurement of 
$\bar\nu_e\rightarrow \bar\nu_\tau$ oscillation
 beyond the current limit
set by the Chooz experiment.

We remark that if the modulus of the matrix elements (other than the four
we used as input) is experimentally determined, we can 
predict the CP violation phase $\phi$ as,

\begin{equation}
\cos\phi=\frac{1}{2 |U_{e 1}| |U_{e2}||U_{e3}| |U_{\mu 3}||U_{\tau 3}|}
 [(1-|U_{e3}|^2)^2 |U_{\mu 1}|^2 - |U_{e2}|^2 |U_{\tau 3}|^2
-|U_{e1}|^2 |U_{e3}|^2 |U_{\mu 3}|^2].
\end{equation}
  
We do not repeat the discussion about actual CP violation effects given
in the literature [19], but let us quote that the disparity of neutrino
oscillation due to CP violation is given by [21]

\begin{eqnarray}
\Delta P&=& P(\bar\nu_\mu\rightarrow \bar\nu_e)-P(\nu_\mu\rightarrow \nu_e)\cr
         &=& 4J f \leq 0.16 f\sin\phi\ ,
\end{eqnarray}
where $
f=4\sin(\Delta_{12}/2)\sin(\Delta_{32}/2)\sin(\Delta_{31}/2)
$
with $\Delta_{ij}=\Delta m_{ij}^2 L/2E_\nu$ ($L$ is the length of the
baseline, and $E_\nu$ is the neutrino beam energy). 
The upper limit of $J$ corresponds to about 2/3 the value
assumed in the analysis of Arafune et al.[19], but if it takes a
value close to this limit the CP violation effect is probably
visible.

So far we have discussed the case of the LMA solution. The experiments
also allow the LOW solution, which are located close to the bottom of
the MSW triangle, albeit the parameter range allowed by a 90\% confidence 
is narrow, If LOW is the solution, the mixing matrix is

\begin{equation}
U= \left[ \matrix{0.71-0.79 & 0.61-0.71 & <0.16 \cr
                  0.34-0.65 & 0.42-0.70 & 0.57-1/\sqrt 2 \cr
                  0.25-0.58 & 0.32-0.63 & 1/\sqrt 2-0.82 \cr
                                         } \right]\ .
\end{equation}
This matrix is quite similar to the one given for LMA, but the constraint
is slightly tighter because of a smaller allowed region for $U_{e2}$. 
We obtain a 
CP violation $J$ factor similar to (5). 
The observable CP violation effect,
however, contains the mass square difference, and the small
$\Delta m_{12}^2$ of the LOW solution pushes the effect outside
the range feasible with accelerator
experiments. 
We should await the KamLAND experiment [22] for a decisive answer as to the
selection between the two solutions.

\vskip 10mm
\noindent
{\large\bf Acknowledgements}

We thank Tsutomu Yanagida for many valuable discussions over many aspects
of neutrino physics, including the present work.

\newpage
\noindent
{\large \bf References}\par
\vskip 0.3 cm
\ref
 [1] SNO Collaboration: Q. R. Ahmad et al., nucl-ex/0106015,
 Phys. Rev. Lett. in press (2001)

\ref
 [2] Super-Kamiokande Collaboration: S. Fukuda et al. Phys. Rev. Lett.
{\bf 86}, 5651; 5656 (2001)

\ref
 [3] J. N. Bahcall, M. H. Pinsonneault and S. Basu, astro-ph/0010346v2 (2001)

\ref
 [4] J. N. Bahcall, P. I. Krastev and A. Yu. Smirnov, Phys. Rev. D{\bf 58},
096016 (1998);  G. L. Fogli, E. Lisi and D. Montanino, Astropart. Phys.
{\bf 9}, 119 (1998)

\ref
 [5] B. T. Cleveland et al., Astrophys. J. {\bf 496}, 505 (1998)

\ref
 [6] GALLEX Collaboration: W. Hampel et al. Phys. Lett. B{\bf 447},
 127 (1999); SAGE  Collaboration:
 J. N. Abdurashitov et al. Phys. Rev. C{\bf 60}, 055801 (1999) 

\ref
 [7] S. P. Mikheyev and A. Yu. Smirnov, Sov. J. Nucl. Phys. {\bf 42}, 913
  (1985); L. Wolfenstein, Phys. Rev. D{\bf 17}, 2369 (1978)

\ref
 [8] G. L. Fogli, E. Lisi, D. Montanino and A. Palazzo, hep-ph/0106247

\ref
 [9] J. N. Bahcall, M. C. Gonzalez-Garcia and C. Pe$\tilde{\rm n}$a-Garay,
hep-ph/0106258

\ref
 [10] A. Bandyopadhyay et al. hep-ph/0106264

\ref
 [11] Super-Kamiokande Collaboration: Y. Fukuda et al. Phys. Rev. Lett.
{\bf 81}, 1562 (1998); T. Toshito, in Proceedings of the 30th 
International Conference on High Energy Physics, Osaka, 2000, 
ed. C. S. Lim and T. Yamanaka
(World Scientific, Singapore, 2001), Vol. 2, p. 913

\ref
[12] E. Peterson, in Proceedings of the 30th International Conference on High
Energy Physics, Osaka, 2000, ed. C. S. Lim and T. Yamanaka
(World Scientific, Singapore, 2001), Vol. 2, p. 907;
F. Ronga et al. {\it ibid} p.910

\ref
[13] K2K Collaboration: S. H. Ahn et al. Phys. Lett. B{\bf 511}, 178 (2001)

\ref
[14] M. Apollonio et al.,  Phys. Lett. B{\bf 466}, 415 (1999)

\ref
[15] Particle Data Group, D. E. Groom et al., Eur. Phys. J. C{\bf 15}, 1
(2000)

\ref
[16] M. Fukugita, M. Tanimoto and  T. Yanagida, Prog. Theor. Phys. {\bf 89},
        263 (1993)

 \ref
[17] M. Fukugita, M. Tanimoto and  T. Yanagida,  
  Phys. Rev. D{\bf 57}, 4429 (1998)
 
\ref
[18] M. Tanimoto, Phys. Rev. D{\bf 55}, 322 (1997)

\ref
[19] J. Arafune, M. Koike and J. Sato, Phys. Rev. D {\bf 56}, 3093 (1997);
see also J. Arafune and J. Sato, Phys. Rev. D {\bf 55}, 1653 (1997)

\ref
[20] C. Jarlskog,  Phys. Rev. Lett. {\bf 55} 1039 (1985)

 \ref
[21] V. Barger, K. Whisnant, S. Pakvasa and R. J. N. Phillips,
 Phys. Rev. D {\bf 22}, 2718 (1986)
 
 \ref
[22] A. Suzuki, Nucl. Phys. B (Proc. Suppl.) {\bf 77}, 171 (1999)

\end{document}